# How do Viruses Attack Anti-Virus Programs

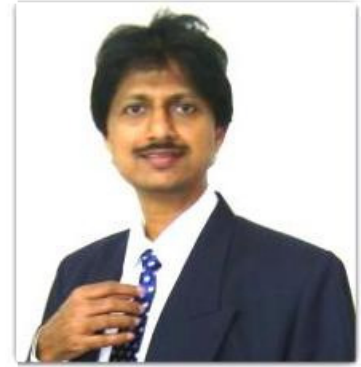

**By- Umakant Mishra, Bangalore, India**

umakant@trizsite.tk, http://umakant.trizsite.tk

**Contents**

## 1. War between viruses and anti-viruses

With the improvement of virus detection technologies the virus creators found it difficult to create viruses that were capable of surviving long. This situation made them to innovate new methods for the survival of their viruses. Some important stealth techniques such as encryption, polymorphism and metamorphism were developed in order to make the viruses capable of escaping conventional scanning methods.

But the war between virus creators and anti-virus creators was far from being over. The anti-virus creators implemented new techniques such as heuristic scanning and emulation techniques which were capable of detecting encrypted polymorphic viruses. Some modern anti-viruses used automatic learning, neural networks, data mining, hidden markov models, static and dynamic heuristics, rootkit heuristics and many other methods to remove almost every virus hidden anywhere in the computer.

## 2. Why is an anti-virus targeted

When the virus creators failed to hide their malware from the advanced detection technologies they came up with various offensive techniques to attack and make the anti-virus program to paralyze. If the anti-virus can be screwed and compromised then the computer can be a safe haven for virus proliferation.

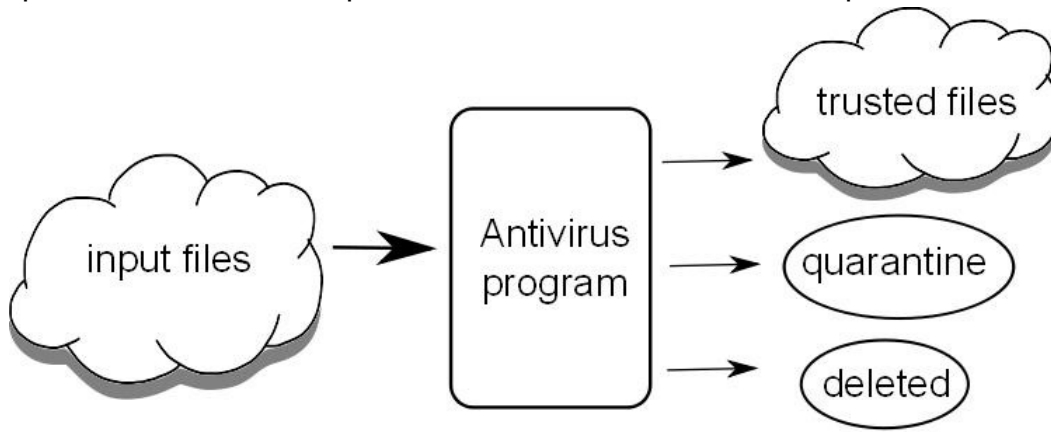

Anti-virus as the security guard for incoming files

The anti-virus software works at a highly trusted level of the operating system. Generally virus creators look for such type of levels to attack the computer system. So the virus creators are always in search of vulnerabilities in operating systems and anti-virus systems which can give them golden opportunity to attack and take control over the computer system.

Traditionally the loopholes of Windows operating system has led to creation of various types of viruses, worms and other malware to attack the Windows based computers. When the loopholes of the operating system are closed by security patches, the loopholes in anti-virus systems open up new avenues for the virus creators to defend their malign creatures.

## 3. History of attacks on anti-viruses

While anti-viruses are built with an intention to protect computers from virus attacks they are not always strong enough to do their intended job. Recently I went to a cybercafe with 4 files in my pendrive. When I opened my pen drive I saw many more files, and the files are increasing and decreasing dynamically. I checked the computer and found that an anti-virus was installed. So I did not think of viruses nor could really understand anything. When I came back home I found 36 files in my pen drive, excluding my original 4 files, the rest 32 are worms, trojans and other viruses. I formatted the drive.

Any vulnerability in the anti-virus becomes an opportunity of the virus creators.
All anti-virus vendors including popular names like McAfee, Symantec, TrendMicro, VBA32, Panda, PC Tools, CA eTrust, ZoneAlarm, AVG, BitDefender, Avast!, Kaspersky, Sophos etc. have faced such kind of experiences.

BusinessWire[#] mentions that about 800 vulnerabilities (in 2008) including system access, DoS (denial of service), privilege escalation, security bypassing etc. were discovered in various anti-virus products. *"The conclusion: contrary to their actual function, the products open the door to attackers, enable them to penetrate company networks and infect them with destructive code. …".* Let's see why the attackers try to exploit such vulnerabilities:

## 4. Why is it easy to attack anti-viruses

Viruses are present in almost every system even though there are anti-viruses installed. If the anti-virus is not reliable or not updated or compromised then the viruses can breed and grow safely in the system. Besides every anti-virus , however good that may be, leads to some extent of false positives and false negatives. Our faith on the anti-virus system often makes us more careless about hygienic habits which increases the possibility of infection. Let's analyze the reasons why do the anti-viruses are targeted by the viruses.

- ⇨ Complexity of testing AV product- every anti-virus is bundled with various types of scanning and detection techniques with lots of heuristics and other methods. This makes the scanning engine extremely complex and creates great difficulty for the quality control staff to ensure a proper quality check before releasing the product. Minor imperfections in the detection algorithm not only leads to false positives and false negatives but also leads to vulnerabilities for viral attacks.

- ⇨ Insufficient time for testing AV product- One of the weaknesses of any anti-virus product is that it goes through frequent updates and modifications. With the development of new techniques and algorithms most commercial anti-virus companies upgrade their software product almost in every quarter. This has a serious impact on in-house testing and debugging as the developer never gets enough time to test the new components implemented in every release, thereby leaving some unfixed bugs in the product.

- Usually people have tremendous faith on anti-virus programs. Even if the anti-virus program malfunctions or behaves in slightly unnatural way people never bother about such anomalies. (For example, my AVG anti-virus gives "avgmfapx.exe error…" message every 3-4 days. But I hardly bother to read the details and find any remedy). This mindset of the users helps the virus creators to exploit the vulnerabilities of anti-virus with greater ease.

## 5. Methods of fooling and attacking the anti-virus

As we discussed above, the vulnerabilities of anti-virus programs are exploited by the viruses and hackers. Besides in some cases there are bugs in anti-virus programs that give great opportunities to the attackers. Let's see some methods of cheating and attacking the anti-virus by different viruses.

- Memory resident viruses- one of the oldest methods of defeating the anti-virus is by using memory resident techniques. The memory resident viruses remain resident in the memory and try to deceive the anti-virus program. They hook to the Int 13h and keep on fooling the anti-virus program by paralyzing the crucial functions of the anti-virus program.

- Boot sector replacement- boot sector viruses may modify the actual MBR and create a façade MBR at a location other than the actual location of the MBR and fill the façade with the content of the original MBR. When the anti-virus checks the MBR to determine its integrity the comparison is performed on the façade MBR instead of the modified MBR. This method hides the infection from integrity checker.

- One of the weaknesses of anti-virus programs is that they cannot scan the compressed files. Hence the anti-virus needs to decompress the compressed files and data before scanning. But the attackers sometimes create specially made compressed files using nested compression or using infinite loops to fool the anti-virus. When the anti-virus tries to decompress the file it falls into an infinite loop which causes high CPU usage and leads to Denial of Service (DoS).

   In order to avoid this problem some anti-virus programs use a time-out method to stop protracted scanning and get out of the infinite loop. However it is difficult to decide whether the compressed file is a genuine file or a virus made special file to fool the anti-virus. This situation leads to the following contradiction.

> *Contradiction*
> 
> *If a protracted scanning is not timed out, then the anti-virus system will continue using system resources and impacting the system performance. But using a time-out may inappropriately terminate a genuine scanning operation in a slow or overloaded computer. We want to terminate the prolonged scanning operations to save system resources but don't want to terminate a prolonged scanning in a genuinely slow or stressed system.*

⇨ Similarly, in some other cases the attacker uses specially compressed files that require huge amount of disk space. For instance, a small zip file may consume as much as 4GB of disk space and make the hard disk full leading to "insufficient disk space" error.

⇨ Similarly the encrypted and password protected files cannot be checked for viruses because they cannot be seen from outside. Many viruses and hackers pack their malware code in zip files, encrypted files, password protected files etc. and send them as email attachments to potential victims (sometimes even with the decryption password). As in most cases the scanner is not able to scan the content of the encrypted files, the malware code remains in the computer for long time.

⇨ Sometimes rootkits are used to gain administrative-level control over a computer system. A rootkit provides the modified versions of the original utilities used for administering a computer. A rootkit infected computer do all its usual functions along with the special jobs programmed inside the rootkit. Often the rootkits are used to avoid detection of backdoors and specific virus infections.

⇨ Using bait files or exploit files – the exploit files are built specifically keeping the vulnerability of specific anti-viruses in mind. The attackers may put the exploit files on their websites and convince victims to download the exploit files by using social engineering techniques. Alternatively the exploit file may get downloaded without user's explicit permission. When the anti-virus of the user computer scans the exploit file, the anti-virus engine is compromised.

⇨ Local privilege escalation problems- when the anti-virus software gives "Full control" permission to "everyone" group, anyone can modify installed files. As every anti-virus software has to run some system services, the attackers may replace an installed service file with a malicious Trojan or rootkit which can later run with "system" privileges.

- ⇨ Some ActiveX controls may have design errors which include insecure methods. Besides ActiveX controls often lead to memory corruption. Attackers will pass very long strings as parameters to the vulnerable ActiveX controls in order to cause a memory corruption.

- ⇨ Some vulnerabilities are caused by misutilising the SERVUCE_CHANGE_CONFIG permission. The attackers may use sc.exe to stop an anti-virus service and start it again after making modifications in the anti-virus program or environment.

- ⇨ Loopholes at the mail server- some mail servers are configured to scan mails by anti-virus scan engines. Such situations create opportunities for the attackers to exploit the loopholes of the anti-virus engines. The attacker will send virus emails to randomly collected addresses. When the engine will scan the emails the engine and the mail server will be compromised.

- ⇨ Loopholes at the mail client- individual users may be attacked by sending virus emails which will take advantage of the anti-virus engine vulnerability. The victim's computer will be compromised when the email is scanned by the anti-virus installed on his computer.

- ⇨ Generally the anti-virus stores the previously scanned data in an AV state database in order to avoid repetitive scanning of the same files in subsequent scanning sessions. Some malware target to attack these AV state database and try to change the status of files to "virus free" in order to avoid scanning of infected files by the anti-virus.

- ⇨ The modern viruses use many anti-anti-virus techniques to escape detection. There are anti-emulation techniques to avoid CPU emulation and anti-heuristic techniques to avoid heuristic scanning. These techniques fool the anti-virus systems and make their effort go waste.

- ⇨ Attacks on signature scanning- some viruses attempt to alter the virus scanning executables and virus signature databases so that they cannot detect the viruses. However with the development of security mechanism these operations have been extremely difficult.

- ⇨ Attacks on integrity checkers- there are some viruses which target to integrity checkers and infect only on intentional file modification. When the user allows the changes to the executable file, the infections are also allowed and added to the database. Besides, some viruses use "stealth" techniques to bypass file reads thus fooling the integrity checkers which believe that the checksums remain the same.

- Some viruses target to manipulate the integrity database which stores the checksums or hash strings of the original files. They try to delete or modify the integrity database in order to confuse the integrity checker and avoid detection. If the integrity checker recreates the deleted integrity database, it has to calculate the checksums of the infected files.

- Fooling the CPU emulator- some viruses try to fool the CPU emulator. When the encrypted virus decrypt its body it shows its body only part by part so that only a small part of their decrypted body is visible to the scanner at any point of time. As the decrypted portion is small the scanner cannot find any virus signatures in the decrypted portion.

- Attacking memory scanning- Some viruses may work simultaneously on multiple processes. In other cases multiple copies of a worm may run watching each other. These situations are difficult for the anti-virus to tackle.

## 6. How to protect anti-viruses from viral attacks

As the anti-viruses run in a trusted kernel level any loophole in the anti-virus program can enable attackers to take full control over the computer system and steal data or do serious damages. Hence the anti-virus engines must be developed with proper security in mind. The ant-virus should be able to any type of specially created executable files, compression packages or documents that are intentionally created to exploit the anti-virus's weakness.

It is necessary for an anti-virus to detect and destroy the malware before its own files are detected and destroyed by the malware. Is it necessary to install another software such as firewall, Intrusion detection System or something else to protect the anti-virus? We will discuss the methods of protecting anti-virus program from malware attacks in a separate article.